\documentclass[12pt]{iopart}
\usepackage{iopams}  
\usepackage{graphicx}
\begin{document}

\title[Understanding CMB physics using CLASS]{Understanding CMB physics through the exploration of exotic cosmological models: \\a classroom study using CLASS}

\author{T. Rindler-Daller}

\address{Institut f\"ur Astrophysik, Universit\"atssternwarte Wien, University of Vienna, \\T\"urkenschanzstr. 17, 1180 Vienna, Austria}
\ead{tanja.rindler-daller@univie.ac.at}
\vspace{10pt}
%\begin{indented}
%\item[]August 2017
%\end{indented}

\begin{abstract}
Every cosmology lecturer these days is confronted with teaching the modern cosmological standard model $\Lambda$CDM, and there are many approaches to do this.
However, the danger is imminent that it is presented to students as something set into stone, merely to be accepted as a fact based on the plenty of evidences we have.
This is even more critical, given that the standard model of cosmology confronts us with entities not yet fully understood, namely a cosmological constant $\Lambda$ and cold dark matter.
In this article, we report on an advanced cosmology course exercise, conducted in computer lab, which was conceived as a means to have students experience first-hand why the 
$\Lambda$CDM model has become so prevalent in the interpretation of modern cosmological data. To this end, we focused on the cosmic microwave background radiation (CMB)
and calculated theoretical temperature and matter power spectra, using the modern Boltzmann code CLASS. By comparing and analyzing the outcome for $\Lambda$CDM, 
as well as for three other exotic cosmological models, the students were able to grasp the impact of cosmological parameters on CMB observables, and also 
to understand some of the complicated CMB physics in a direct way. Our chosen examples are not exhaustive and can be easily modified or expanded, 
so we express the hope that this article will serve as a valuable resource for interested students and lecturers.
\end{abstract}

\vspace{2pc}
\noindent{\it Keywords}: cosmological parameters, cosmic microwave background radiation, expansion history, structure formation
%\maketitle

\section{Introduction}
\label{sec:Intro}
The cosmological standard model $\Lambda$CDM has been confirmed as a 'best-fit model' by many observations of the last two decades, 
notably by measurements of the large-scale structure via galaxy surveys (e.g. \cite{S1,S2}), the cosmic microwave background radiation (e.g. \cite{CMB1,CMB2}), 
and the distance ladder using variable stars and supernovae (e.g. \cite{D1,D2,D3,D4}). These lists of references are by no means exhaustive, but they represent
a sample of very recent constraints. The status of $\Lambda$CDM as a best-fit model concerns mostly gross variables, such as cosmic density parameters, and parameters 
which relate the initial conditions to the present. However, the nature of the two main ingredients of $\Lambda$CDM - the cosmological constant $\Lambda$ and 
cold dark matter (CDM)-, is still unsettled, despite many past and on-going efforts to reveal their detailed characteristics. This is the reason why 
$\Lambda$CDM has been more honestly called the ''current concordance model'' in its early days around the turn of the millenium. However, since then it has morphed more and more 
into our current cosmological standard model.

Teaching the cosmological standard model requires to make reference to many crucial observations like those indicated above, and students are often merely presented with the fact that these observations are nicely fit with $\Lambda$CDM. It is then not obvious to many what it would entail, if models with drastic deviations from $\Lambda$CDM would be considered, instead.
An important case in point is the cosmic microwave background radiation (CMB), whose measurements have become very precise over the last decade thanks to dedicated space and ground-based facilities. Now, a student might ask: how would the familiar CMB temperature power spectrum look like, if there were no CDM, or no $\Lambda$ in the Universe?

As part of her lecture course on \textit{Cosmological structure formation: theoretical foundations and modern applications}\footnote{in German: \textit{Kosmologische Strukturbildung: theoretische Grundlagen und moderne Anwendungen}}, taught in the winter semester of 2018/19 at the University of Vienna, the author devised a laboratory exercise, in which students learned to use the modern Boltzmann code CLASS, in order to investigate wildly different models and their impact on various cosmological observables. In doing so, the students not only acquire the skill to use modern cosmological software, but they also appreciate first-hand why $\Lambda$CDM has become so prevalent and important in the interpretation of modern observations. 

There is yet another reason why this exercise was conceived. The physics of the CMB is relatively complicated, and students may feel a gap between 
learning some of the physical principles and relating them to observations, e.g. to the form of the CMB temperature power spectrum. 
In particular, the understanding of various interactions between baryons and photons, together with the dark matter, is a prerequisite in the 
interpretation of that power spectrum and other observables. Different assumptions on cosmological parameters imply different density perturbations and structure formation
scenarios. Thus, calculating CMB spectra for models which are very different from $\Lambda$CDM 
helps, in turn, to gain a better understanding of the impacts of baryons, dark matter, or $\Lambda$ on the CMB.

In fact, before modern observations were available, early works have attempted to predict the detailed form of the CMB spectra for models of varying baryon or CDM content,
from first principles using analytic calculations (see e.g. \cite{HS95, HS96, Mukh}). It is advisable to study such papers, individually or in class, 
in order to gain a thorough understanding of the topic.
However, it turns out that our approach works well, particularly in cases when the time in lecture is limited, or when the background of students is very 
diverse\footnote{In my experience, beginning graduate students often have had only limited or no exposure to advanced cosmology, before entering this field.}.
CLASS is open source software, can be easily used in a classroom environment, and gives quick results which can be discussed and analyzed. 
In this sense, our exercise could even be useful in undergraduate courses, and the degree of analysis is merely subject to the background of the participants.
Let us also note that there is a variety of web-based applets available which allow the user to change parameters at the touch of a button or slide bar and see live how the CMB spectra
change. It is needless to say that such an approach is no proper replacement, for it seems like a ``black box'', and does nothing in teaching students to use real 
scientific software which is standard in the field.

The exotic models investigated and described below were chosen purely for the sake of didactic usefulness and \textit{not} because they are supposed to describe reality. 
They do not! The lecture course covered many topics on structure formation, so due to time constraints we could not study more models
in CLASS, than the ones described in this article. In addition, various parameter degeneracies made it necessary to focus on certain changes and their impact, as opposed to sampling an exhaustive number of illustrative models.
With this in mind, I like to emphasize that many more possibilities could be easily studied with CLASS, in general. 
The author hopes that this article may help inspire similar experiments, whether performed in class or as individual student.
However, this paper is not the place to present a tutorial on CMB physics, nor to include detailed derivations of the fundamental euqations; 
we refer to the above cited analytic papers, as well as to
more recent reviews, e.g. \cite{Sugi, Staggs} for more background. 

This article is organized as follows: in Section \ref{sec:class}, we describe our use of the CLASS code. In Section \ref{sec:models}, 
we present our chosen cosmological models. Section \ref{sec:insight} concerns the questions we have studied 
and a discussion of some of the most important insights the students gained upon performing this exercise. Section \ref{sec:summary} presents a short summary. 
In the Appendix, we briefly discuss the two versions of $\Lambda$CDM with which we were concerned in our study.

\section{Using the CLASS code}
\label{sec:class}

During the advent of modern CMB observations since the nineties, the community recognized the necessity to have powerful codes, which are able to calculate in detail 
the background evolution and the linear growth of structure formation up to and beyond the time of decoupling, after which photons started to stream freely in the Universe. 
The redshifted "light" from that time - the redshift of the surface of last scattering -  is now seen as the CMB. These codes allow to probe many different physical effects and their impact onto the CMB. 
In turn, these theoretical models can be compared to actual CMB data. There are several such codes available. I chose to use CLASS for the purpose of my course,
because I have acquired experience with CLASS in my own scientific work and, more importantly, the modular structure of CLASS makes it very easy to learn and to use.
CLASS has been developed by Julien Lesgourgues \cite{CLASS}, and there is a stream of further methodology papers \cite{CLASS2, CLASS3, CLASS4}. CLASS is written in C; the code is well documented and can be downloaded for free at \verb"http://class-code.net/".
We used CLASS version 2.7.1, dated from September 2018. The students downloaded CLASS, made sure it would compile (by adjusting the Makefile, if needed) and first 
ran a $\Lambda$CDM model. The $\Lambda$CDM model served as a basic reference to compare with the other exotic models to be described in the next section.
Each version of CLASS usually comes with an input file tailored to produce a $\Lambda$CDM model, as currently favoured. In the lab, we encountered two versions, 
an older $\Lambda$CDM model from the first releases called \verb"lcdm.ini", and the newest one which comes with CLASS 2.7.1, called \verb"base_2018_plikHM_TTTEEE_lowl_lowE_lensing.ini".
The latter is based on the parameters of Case 2.17 of
\verb"https://wiki.cosmos.esa.int/planck-legacy-archive"\\
\verb"/images/b/be/Baseline_params_table_2018_68pc.pdf", as part of the latest release of Planck data \cite{CMB1}. In the Appendix, we discuss the differences.

Compared to the exotic models, the differences between these two $\Lambda$CDM models (old and new) are very marginal. Of course, small deviations or tensions between various
observations can make a large difference in the theoretical interpretation and the respective comparison to data in the everyday life of cosmologists! But, obviously, this is not the 
focus of our study here.
Yet, as a matter of fact, codes like CLASS have been devised in the era of precision cosmology, and it is not a matter of course to find out that the code 
still works well, even if we feed it with very exotic models, with parameters wildly different to the standard model. 
In my opinion, it is fair to say that this speaks for the robustness and reliability of CLASS.  
We emphasize again that the purpose of this exercise consisted not only in learning to use modern scientific software,
but mostly to grasp the impacts of baryons, CDM and $\Lambda$ onto observables related to the CMB. 
We studied the cosmological standard model, as well as some exotic models as examples and compared the respective output of these models. We did not work with CMB data itself.

\section{Models}
\label{sec:models}

The standard cosmological model $\Lambda$CDM is based upon general relativity and the cosmological principle of homogeneity and isotropy of the background universe.
We adopt this premise for our exotic models as well, i.e. the background geometry is in each case described by a Friedmann-Lemaitre-Robertson-Walker (FLRW) metric
of constant curvature, which in spherical polar coordinates $(r,\theta,\phi)$ is written as
\begin{equation} \label{flrw}
ds^2 = dt^2 - a^2(t)\left[\frac{dr^2}{1-Kr^2} + r^2 (d\theta^2+\sin^2\theta d\phi^2) \right],
\end{equation}
where the curvature distinguishes spherical ($K = 1$), flat ($K=0$), and hyperbolic ($K=-1$) geometries, $t$ is the cosmic time and $a$ is the scale factor. 
The calculation of linear perturbations on top of this metric are also carried out in a standard way, for all models considered. 
We have not changed \textit{any} equations in CLASS, whether they relate to the dynamics or thermodynamics.

The evolution of the background universe, notably the expansion history, is determined by the Friedmann equation, which is nothing but the time-time component of 
the Einstein field equation\footnote{$c$ is the speed of light, $G$ is the gravitational constant}, using the FLRW metric: 
\begin{equation}
H^2(a) = \frac{8\pi G}{3}\sum_i \rho_i(a) + \frac{\Lambda c^2}{3} - \frac{K c^2}{a^2}.
\end{equation}
The expansion rate or Hubble\footnote{I use the mathematics convention, where the symbol ``:='' indicates a definition.} parameter, $H := (da/dt)/a$, is determined by the sum of the energy densities of relativistic and non-relativistic cosmic components 
$\rho_i(a)$, the cosmological constant $\Lambda$ (if present), and the curvature term. Except for $\Lambda$, all these contributions evolve as a function of $t$,
or $a$, respectively.  By convention, the present is set at $a=1$. In the standard model, the energy densities of dark matter $\rho_{\rm cdm}$ and baryons $\rho_{b}$ evolve
like non-relativistic matter as of a very early time, decaying as $\propto a^{-3}$ thereafter, and we also adopt this behaviour for our exotic models. Photons and neutrinos (``the radiation'')
evolve like $\rho_{\rm rad} \propto a^{-4}$. The cosmological constant term can be parameterized as an effective energy density, according to $\rho_{\Lambda} = \Lambda c^2/(8\pi G)$.
We do not introduce any further cosmic components in our exotic models, but only study what happens, if we left out a part, or changed the amount of the standard inventory.
It is customary to introduce and work with fractional energy density contributions, the so-called density 
parameters, $\Omega_i := \rho_i/\rho_{c}$ with the
critical density defined as $\rho_{c} := 3H^2/(8\pi G)$, i.e. $\Omega_{\rm cdm} := 8\pi G \rho_{\rm cdm}/(3H^2)$,
$\Omega_b := 8\pi G \rho_b/(3H^2)$,  $\Omega_{\rm rad} := 8\pi G \rho_{\rm rad}/(3H^2)$,  $\Omega_{\Lambda} := \Lambda c^2/(3H^2)$,  $\Omega_k := -Kc^2/(a^2 H^2)$.
Then, the above Friedmann equation takes the form
\begin{equation} \label{closure}
 1 = \sum_i \Omega_i + \Omega_{\Lambda} + \Omega_k,
\end{equation}
which is sometimes called a ``closure condition''; in flat geometries $\Omega_k = 0$. Of course, the latter is favoured by the standard cosmological model $\Lambda$CDM.  
In what follows, we include a subscript $0$ to designate the present-day values of cosmological parameters.  It is actually these present-day values which are input to our CLASS
calculations.

The students were advised to study the following models using CLASS:

\textit{Except for one example}, all models considered were flat, $\Omega_{k} = 0$, which implies some restrictions on the kind of exotic models considered. 
Also, we kept many default parameters as they are informed from our experience with best-fit $\Lambda$CDM. However, we changed some of the
most important cosmological parameters in our study, in order to force drastic deviations from $\Lambda$CDM, and yet be in accordance
with equ.(\ref{closure}). In some cases, we had to adjust certain other parameters in order to run the calculations successfully.
Except for small differences in the present-day value of the Hubble parameter $H_0$ -or the reduced Hubble parameter $h := (H_0/100)$ s Mpc$/$km-
between the old and the current version of $\Lambda$CDM, we did not experiment with $H_0$.
Also, we did not experiment with some other parameters: we used the default values of the number of relativistic degrees of freedom $N_{\rm eff}$, as well as the 
amplitude and index of the primordial scalar spectrum, $A_s$ and $n_s$, respectively,
as they were set in the respective input files.  The present energy density of radiation can be determined from the mean CMB temperature; it is about
$\Omega_{\rm rad,0} \simeq 5.05 \times 10^{-5}$ and is basically the same for each model.
Details to the most important input parameters can be found in Table \ref{table1}, where the input file for the standard model $\Lambda$CDM comes already with CLASS (details below), while
the input files for our exotic Models 2-4 were adapted using \verb"explanatory.ini" which comes with
CLASS release 2.7.1. 

For all runs with CLASS we chose adiabatic initial conditions and no lensing (although the input file of $\Lambda$CDM (2018) has
lensing included, but that makes basically no difference to the results of our study).  Adiabatic initial conditions are predicted by the most successful and simple
inflationary models which are based on a single scalar field. In fact, Planck data has constrained non-adiabatic contributions to the CMB temperature power spectrum
to less than a few percent. Thus, for comparison's sake, we picked the same initial conditions for all of our models.  
Furthermore, we chose the Newtonian gauge in all of our runs. The choice of gauge may be determined by various purposes, but since we consider one model without CDM, it seemed
practical to pick the Newtonian gauge for all models and it worked fine for each case.

\begin{itemize}
\item the \textbf{cosmological standard model $\Lambda$CDM}: 
In this article, we present results for the most up-to-date best-fit $\Lambda$CDM model from Planck 2018, which comes with CLASS release 2.7.1 (see Sec.\ref{sec:class});
we call it $\Lambda$CDM (2018). However,
in the lab we actually focused on an older, ``less complicated'' version, which was sufficient for the purpose of our study; we call it $\Lambda$CDM (2011), see also 
the Appendix. As can be seen from Table 1, $\Omega_{b,0}h^2 = 0.0223828$ and $\Omega_{\rm cdm,0}h^2 = 0.1201075$. Using the respective Hubble constant, these values
correspond to $\Omega_{b,0} = 0.0493868$ and $\Omega_{\rm cdm,0} = 0.2650127$, i.e. the present total energy density consists of roughly $5\%$ baryons and $26\%$ CDM.
Using the closure condition in equ.(\ref{closure}), CLASS derives the present energy density of $\Lambda$; using the above values we have $\Omega_{\Lambda,0} = 0.6856005$.
Note that the contribution of the radiation at the present is so small, that it is usually neglected.
$\Lambda$CDM served as our reference model to which the other, exotic models were compared. 
\item \textbf{Model 2: no CDM component}: here, we set $\Omega_{\rm cdm,0} = 0$. This is possible in Newtonian gauge; it was not possible to set $\Omega_{\rm cdm,0}$ to zero
in previous versions of CLASS which were limited to synchronous gauge. We leave the baryon content as provided in \verb"explanatory.ini", i.e.
$\Omega_{b,0}h^2 = 0.022032$. Using the respective number for the Hubble constant, we get $\Omega_{b,0} = 0.048275$, i.e. both values are very close to those
for $\Lambda$CDM (2018). 
However, we pick $\Omega_{k,0} = -0.01$ in this model and fulfill the closure condition by forcing
$\Omega_{\Lambda,0} = 0.961725$. Thus, Model 2 is very much $\Lambda$-dominated at the present, 
hence reminiscent of the de Sitter model. Also, it may mimick a MOND (MOdified Newtonian Dynamics) universe, for we have only baryonic matter in this model.
\item \textbf{Model 3: no $\Lambda$ component}: CLASS requires the specification of $\Omega_{\Lambda,0}$ or the density parameter of another dark energy component, 
whether it be a fluid or a scalar field. If the latter are disregarded, $\Omega_{\Lambda,0}$ is inferred by the code automatically via (\ref{closure}), i.e.
it cannot be set equally to zero per se in the input file. Thus, even if (\ref{closure}) is fulfilled algebraically,
CLASS limitations and rounding errors will produce a non-vanishing $\Omega_{\Lambda,0}$, which is very small, however ($\Omega_{\Lambda}$
stays below about $10^{-5}$ up to the present). Since we pick again the same baryon content as in Model 2, i.e. $\Omega_{b,0} = 0.048275$, close to $\Lambda$CDM, and
a flat geometry, $\Omega_{k,0} = 0$, the closure condition requires us to set the CDM fraction to $\Omega_{\rm cdm,0} = 0.951725$, which we use in the respective input file.
As a result, this model universe is strongly CDM-dominated at the present, and is a good approximation for the Einstein-de Sitter model.
\item \textbf{Model 4: more baryons}. While Models 2 and 3 have basically the same amount of baryons as $\Lambda$CDM, we now want to consider a model with 
substantially more baryons,
but we keep the $\Lambda$ and CDM components, as well. We use the default CDM fraction of \verb"explanatory.ini", which is
$\Omega_{\rm cdm,0}h^2 = 0.12038$, i.e. quite close to the value of $\Lambda$CDM (2018). Using the respective Hubble constant, we have $\Omega_{\rm cdm,0} = 0.263771$.
Now, we set the fraction of baryons to equal roughly the fraction of CDM in standard $\Lambda$CDM, i.e. we set
$\Omega_{b,0}h^2 = 0.12$, corresponding to $\Omega_{b,0} = 0.262938$. CLASS will give an error message, because this model is in 
conflict with Big Bang nucleosynthesis (BBN) constraints. This is a great example, for it shows the students that our cosmological 
standard model is informed by many probes! In order to run this model, we forced it to be in conflict with BBN by increasing the Helium abundance to 28\%, beyond the
allowed BBN value.
Again, as we keep a flat geometry in this model, the amount of $\Lambda$ is correspondingly smaller than in $\Lambda$CDM, 
namely $\Omega_{\Lambda,0} = 0.473291$. While such a model is ruled out, not only by BBN, as we will see, it would be an interesting gedankenexperiment to picture such a universe.
For one thing, since the baryons in this case are not sub-dominant any longer, we would not call it ``$\Lambda$CDM''.
\end{itemize}

For each of the considered models, we focused on basically three types of observables:
\begin{enumerate}
\item The run of density parameters $\Omega_i$ as a function of scale factor $a$. This information can be found in output files, ending in \verb"_background.dat".
\item The temperature power spectrum of the CMB as a function of mode number $l$. This information can be found in output files, ending in \verb"_cl.dat".
\item The matter power spectrum as a function of wavenumber $k$.
This information can be found in output files, ending in \verb"_pk.dat".
\end{enumerate} 
Output (i) concerns the evolution of the (unperturbed) background energy densities, i.e. basically the expansion history of our model universes, while
output (ii) and (iii) concern the spectrum of perturbations in the photon component and in the total matter component (CDM plus baryons), respectively.

After the calculations were finished, the next task consisted in making plots of these quantities for each of the considered models,
to note all basic properties (age of the model universes, redshift of matter-radiation equality, redshift of recombination) and to compare the results among each other as a consistency check.
Then, the students were asked to study and interpret their results, compare the different models with each other, and summarize the procedure, their 
results and their interpretation in a written report (a total of 19 students completed successfully the lecture course). In the next section, I will summarize some of the key insights the students could draw by going through this particular lab exercise.

Fig.\ref{figure1}-\ref{figure3} show plots of the considered observables for the different models, as labelled above. Table \ref{table1} shows some key 
input parameters, while Table \ref{table2} shows some basic output quantities, as calculated by CLASS. Similar figures and tables were also produced by 
the students and collected in their reports.

\begin{table}
\caption{\label{table1} Important input parameters of our models$^{a}$ as used in CLASS; some of the parameters are derived in the course of the runs.}
\footnotesize
\begin{tabular}{@{}lllllll}
\br
General & CLASS & $\Lambda$CDM (2011)$^{b}$ & $\Lambda$CDM (2018)  & Model 2  & Model 3 & Model 4 \\
\mr
$H_0$ & \verb"H0" & - & 67.32117 & - & - & - \\
$h$ & \verb"h" & 0.7 & - & 0.67556 & 0.67556 & 0.67556 \\
$\Omega_{b,0}$ & \verb"Omega_b" & 0.05 & - & - & - & - \\
$\Omega_{b,0} h^2$ & \verb"omega_b" & - & 0.0223828 & 0.022032 & 0.022032 & 0.12 \\
$\Omega_{\rm cdm,0}$ & \verb"Omega_cdm" & 0.25 & - & - & 0.951725 & - \\
$\Omega_{\rm cdm,0}h^2$ & \verb"omega_cdm" & - & 0.1201075 & 0.0 & - & 0.12038 \\
$\Omega_{\Lambda,0}$ & \verb"Omega_Lambda" & derived & derived & derived & derived & derived \\ 
$\Omega_{k,0}$ & \verb"Omega_k" & 0.0 & 0.0 & -0.01 & 0.0 & 0.0 \\
Helium fraction & \verb"YHe" & 0.25 & 0.2454006 & BBN & BBN & 0.28 \\
$T_{\rm CMB}$ & \verb"T_cmb" & 2.726 & 2.7255(?) & 2.7255 & 2.7255 & 2.7255\\
$N_{\rm eff}$ & \verb"N_ur" & 3.04 & 2.03066 & 3.046 & 3.046 & 3.046 \\
$A_{s}$ & \verb"A_s" & 2.3e-9 & 2.100549e-9 & 2.215e-9 & 2.215e-9 & 2.215e-9 \\
$n_{s}$ & \verb"n_s" & 1.0 & 0.9660499 & 0.9619 & 0.9619 & 0.9619 \\
\br
\end{tabular}\\
$^{a}$Model 2: no CDM, Model 3: no $\Lambda$, Model 4: more baryons; $^{b}$see the Appendix for a discussion of the different $\Lambda$CDM models. Note for $\Lambda$CDM (2018):
we could not pinpoint the exact value for the CMB temperature $T_{\rm CMB}$, but it is most likely just the same as in \verb"explanatory.ini", i.e. the default input file
which was adapted for our Models 2-4. 
Also, note that $\Lambda$CDM (2018) contains
one massive neutrino species.
\end{table}
\normalsize

\section{Questions and results}
\label{sec:insight}

\subsection{Background evolution}

Perturbation spectra like the CMB and matter power spectra are notoriously difficult to interpret. Before we do that, it is highly advisable to
understand the expansion history and the time evolution of the energy densities $\Omega_i$ of the different cosmic components. So, we first focused on 
the interpretation of those. Let us note again that we present the evolution as a function of the scale factor of the Universe. While the relationship between scale
factor $a$ and corresponding redshift $z$ is straightforward, $a_{\rm{now}}/a_{\rm{then}} = 1+z$ where at the present time we will set $a_{\rm{now}}=1$ by convention, 
the relationship between scale factor and cosmic time depends upon the cosmological model. It is thus customary to present the cosmological evolution as a function of
$a$ or $z$. Also, I will often follow the cosmologist's parlance and talk about ``time'' when we actually quote scale factors or redshifts. 

Some important questions at hand with respect to the background evolution of the models can be summarized as follows:
\begin{enumerate}
\item What is the age of the model universe? 
\item Which relativistic and non-relativistic cosmic components are present and when do they dominate the total energy density of that universe? 
\item What are benchmark times, like redshift of matter-radiation-equality $z_{\rm {eq}}$ and redshift of recombination $z_{\rm {rec}}$? 
How do they change for the different models?
\end{enumerate}

\begin{figure*}
\begin{minipage}{0.5\linewidth}
     \centering
     \includegraphics[width=8.1cm]{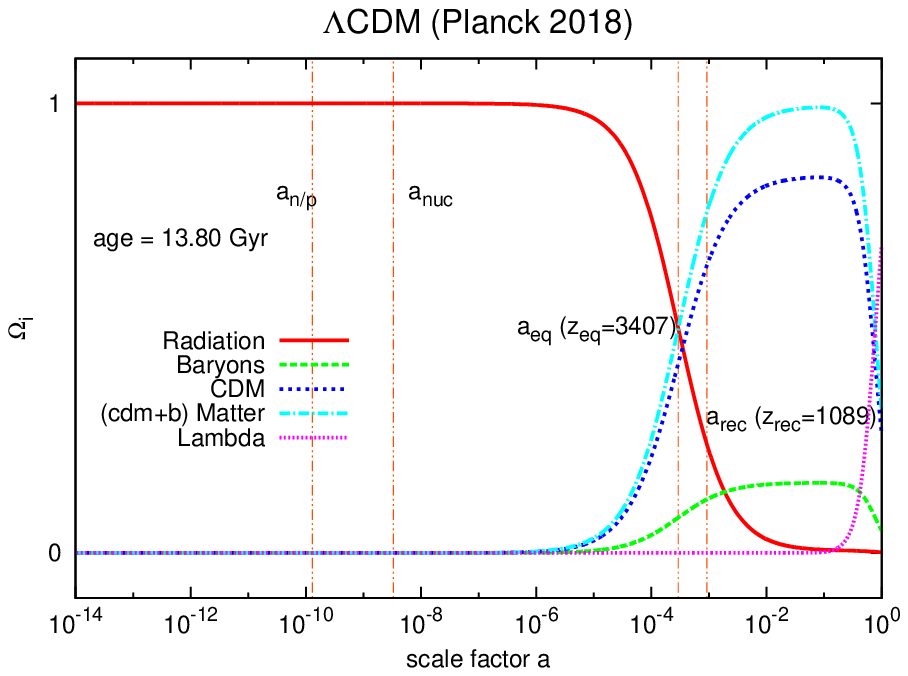}
     \vspace{0.05cm}
    \end{minipage}
    \begin{minipage}{0.5\linewidth}
      \centering\includegraphics[width=8.1cm]{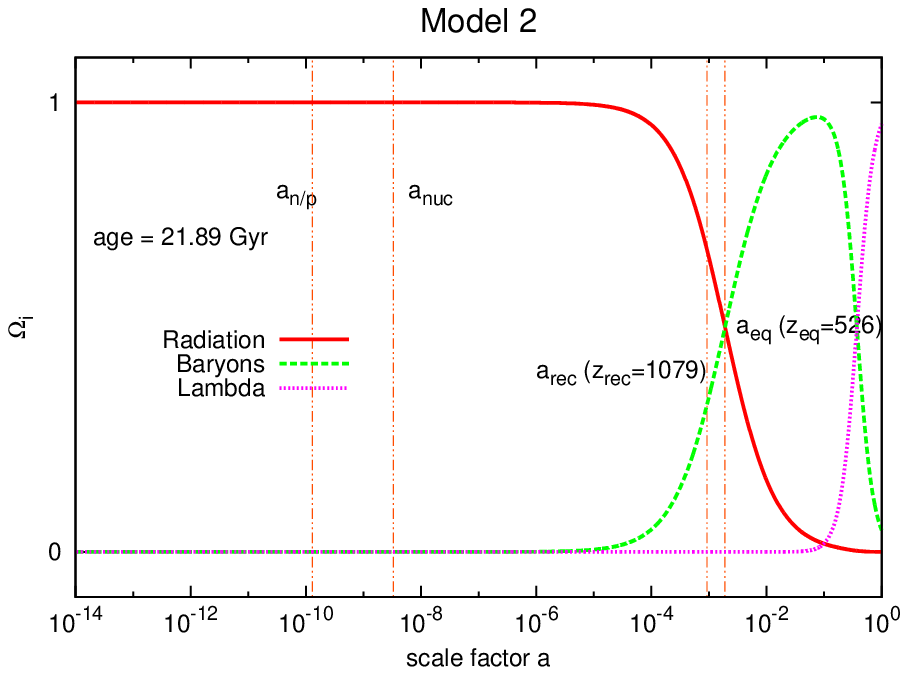}
     \hspace{0.05cm}
    \end{minipage}
 \begin{minipage}{0.5\linewidth}
     \includegraphics[width=8.1cm]{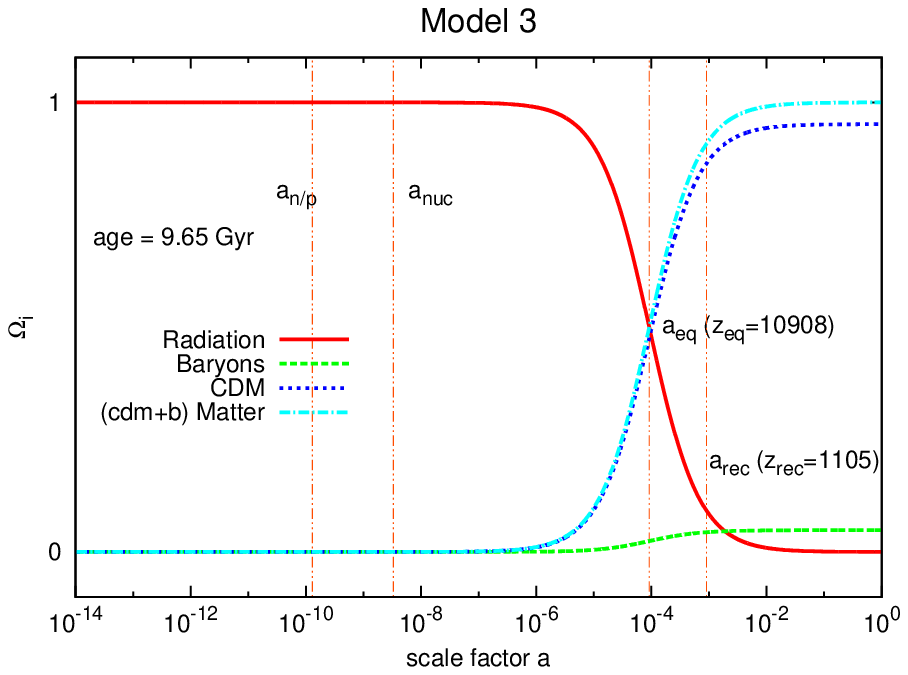}
     \centering
     \vspace{0.05cm}
    \end{minipage}%
    \begin{minipage}{0.5\linewidth}
      \centering\includegraphics[width=8.1cm]{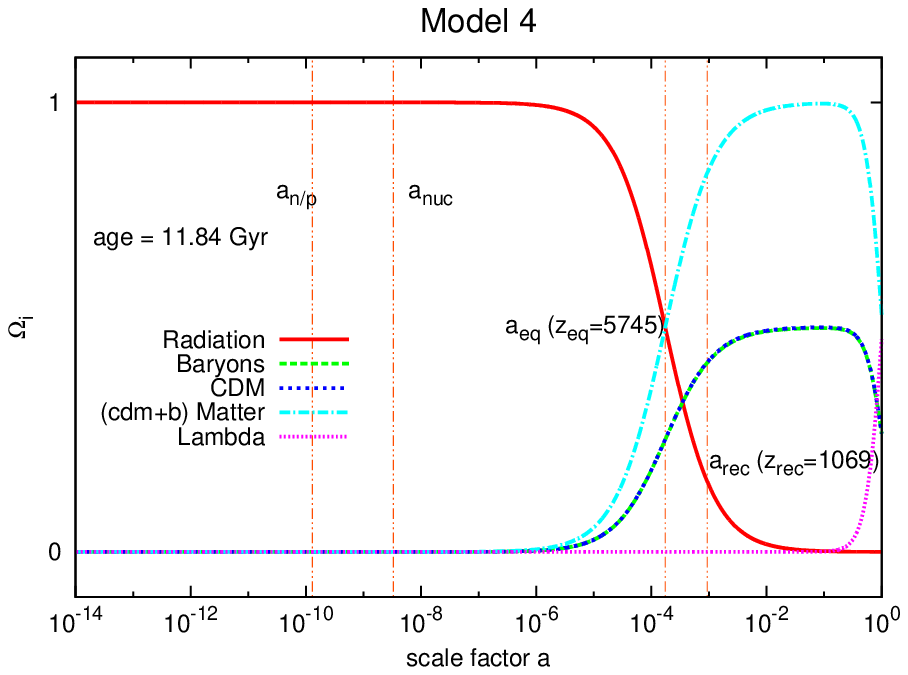}
     \hspace{0.05cm}
    \end{minipage}
 \caption{Evolution of density parameters $\Omega_i$ of all cosmic components in $\Lambda$CDM (top left), in Model 2 'no CDM' (top right), 
 in Model 3 'no $\Lambda$' (bottom left), in Model 4 'more baryons' (bottom right). For each model, we indicate the respective redshift of matter-radiation equality $z_{\rm eq}$
 and redshift of recombination $z_{\rm rec}$. The epoch of BBN is bracketed between $a_{n/p}$ and $a_{\rm nuc}$, and is the same for each model.  
 The age of each model universe is also indicated. More explanations can be found in the main text.}
 \label{figure1}
\end{figure*}

Fig.\ref{figure1} shows plots of $\Omega_i$ versus $a$ for all models considered. I indicated certain important benchmark points: 
the time of BBN, which we bracket by the moments of neutron-proton freeze-out at $a_{n/p} = 1.3011\times 10^{-10}$ and first nuclei production 
around $a_{\rm nuc} = 3.3\times 10^{-9}$. They are the same for each model. Also, the redshifts of matter-radiation equality $z_{\rm eq}$ and recombination $z_{\rm rec}$,
respectively,
are indicated for each case.

Fig.\ref{figure1}, top left, shows the standard model. Obviously, in $\Lambda$CDM radiation (photons plus neutrinos) dominates in the early Universe, 
followed by matter-domination (which is basically CDM-domination), and finally $\Lambda$-domination at the present where $a=1$. We have become 
familiar with this picture over the years. Yet, it is surprisingly absent in most cosmology textbooks, and many
students were already excited by producing that first plot of the overall evolution of $\Lambda$CDM. Since $\Omega_{\rm rad,0} \simeq 5.05 \times 10^{-5}$ 
is basically the same for each model (because the mean CMB temperature $\bar T_{\rm CMB}$ is basically the same),
radiation dominates early, but becomes very sub-dominant before the present, in each case. The total amount of matter differs from case to case, where Model 2 lacks CDM altogether.  This changes the time of 
matter-radiation equality, i.e. the time when the combined energy densities of all relativistic components equal the combined energy densities of all non-relativistic components,
\begin{equation}
 \sum_i \rho_{i, \rm relativistic}(a_{\rm eq}) = \sum_i \rho_{i, \rm non-relativistic}(a_{\rm eq}).
\end{equation}
As in the standard model, the relativistic components are photons and neutrinos, while the non-relativistic components are baryons and CDM (if present) 
(their sum called ``Matter'' in Fig.\ref{figure1}). The age of the $\Lambda$CDM (2018) model is 13.80 Gyrs and the redshift
of matter-radiation equality is at $z_{\rm eq} = 3407$.
Smaller amounts of non-relativistic matter lead to a later time of equality, i.e. lower $z_{\rm eq}$, while higher amounts shift $z_{\rm eq}$ backwards in time. 
The two extremes are given on the one hand by Model 2 without CDM, which has a very low value of $z_{\rm eq} = 526$.
This has been noted early on as a characteristic 
feature -and problem- for a MOND universe. Here, matter-domination sets in very late, and the matter is only composed of baryons, see Fig.\ref{figure1}, top right.
Furthermore, $\Lambda$ starts to dominate earlier than in $\Lambda$CDM, due to the low matter content. In such a universe with its age of 21.89 Gyrs, we
would find ourselves already in a strongly $\Lambda$-dominated epoch.

On the other extreme, we have the CDM-dominated Model 3 without $\Lambda$ with early matter-radiation equality at $z_{\rm eq} = 10908$.
As of then, we have a prolonged epoch of matter-domination, strongly CDM-dominated, right up to the present, 
see Fig.\ref{figure1}, bottom left. Its age is 9.65 Gyrs, close to Einstein-de Sitter, as expected.

On the contrary, Model 4 has all cosmic components available like $\Lambda$CDM, but the amount of baryons is
significantly boosted, which shifts the time of equality correspondingly backwards to $z_{\rm eq} = 5745$. While the sequence of cosmological epochs is close to $\Lambda$CDM, here both matter components contribute roughly equally to the total
matter in the matter-dominated epoch (the curves for CDM and baryons lie almost on top of each other), see Fig.\ref{figure1}, bottom right.
Since the baryon fraction has been boosted at the expense of $\Lambda$, we can also see that the time of matter-$\Lambda$ equality is very close to the present,
and the age of 11.84 Gyrs is lower,
compared to $\Lambda$CDM. 

Indeed, the age of both, Model 3 and 4, is in contradiction with the oldest stars and galaxies we know, so they would be already ruled out on these grounds.

Now let us look at Table \ref{table2} to compare $z_{\rm rec}$, the redshift of recombination which is very close to the redshift of last scattering,
and the redshift $z_{\rm bd}$, which is defined as the time at which the baryons are released from
the Compton drag of the photons. The various numbers for $z_{\rm rec}$ differ from the one of 
$\Lambda$CDM (2018) by less than two percent. The same applies to $z_{\rm bd}$, except for Model 4 where the deviation 
to $\Lambda$CDM (2018) is a little less than ten percent. These numerical results are evidence for the well-known fact that $z_{\rm rec}$ is a very weak
function of the cosmological parameters, $\Omega_{b,0}h^2$ and $\Omega_{m,0}h^2$ where $m$ refers to the total matter content, 
while it depends exponentially on the temperature of the baryon-photon
fluid. The baryon drag epoch ends at a related redshift $z_{\rm bd}$, which depends somewhat stronger on these parameters. Reference \cite{HS96} includes a detailed discussion,
including fitting formulae for these redshifts, which are valid for a broad range of $\Omega_{b,0}h^2$ and $\Omega_{m,0}h^2$. It turns out that $z_{\rm rec}$ and
$z_{\rm bd}$ are approximately equal, if $\Omega_{b,0}h^2 \simeq 0.03$. For $\Omega_{b,0}h^2 < 0.03$, the baryon drag ends after last scattering, i.e. 
$z_{\rm bd} < z_{\rm rec}$. This is the case for all models, except for Model 4 where it is the opposite thanks to its very high baryon content.
In addition, there is an overall delay of photon decoupling in high baryon density models, 
leading to a lower redshift of $z_{\rm rec}$, in general, which is again exemplified by Model 4.

Furthermore, while we normally have that $z_{\rm eq} > z_{\rm rec}$, Model 2 without CDM is the only case when $z_{\rm eq} < z_{\rm rec}$,
thanks to its low matter content, i.e. recombination and the decoupling of photons and baryons happen here in the radiation-dominated epoch, another curiosity of this model.

As a final remark, we note that in all models with $\Lambda$ that contribution is always very sub-dominant in the early Universe, right through the time of recombination, and
becomes only significant for redshifts lower than $z \sim 10$. The ``impact'' of vanishing $\Lambda$ in Model 3 is only related to the fact that we
have kept a flat geometry, so we
pushed $\Omega_{\rm cdm,0}$ to a high value, rendering this model strongly CDM-dominated. This will also change the form of the CMB spectrum, accordingly, 
as described in the next subsection.

\begin{table}
\caption{\label{table2} Some characteristic output quantities as calculated by CLASS for the different models$^{a}$: Age of model universe in Gyr, 
redshift of matter-radiation equality $z_{\rm eq}$, redshift of recombination $z_{\rm rec}$, 
redshift at which baryon drag stops $z_{\rm bd}$. }
\footnotesize
\begin{tabular}{@{}llllll}
\br
 & $\Lambda$CDM (2011)$^{b}$ & $\Lambda$CDM (2018)  & Model 2 & Model 3  & Model 4 \\
\mr
Age [Gyr] & 13.461693 & 13.797336 & 21.888385 & 9.648040 & 11.836051 \\
$z_{\rm eq}$ & 3512.881472 & 3406.907947 & 525.614012 & 10907.532967 & 5744.618924 \\
$z_{\rm rec}$ & 1086.845754 & 1088.798382 & 1078.548663 & 1104.856458 & 1069.203668 \\
$z_{\rm bd}$ & 1064.691266 & 1059.935852 & 1047.082113 & 1075.165535 & 1164.561100 \\
\br
\end{tabular}\\
$^{a}$Model 2: no CDM, Model 3: no $\Lambda$, Model 4: more baryons; $^{b}$see the Appendix for a discussion of the different $\Lambda$CDM models

\end{table}
\normalsize

\subsection{CMB temperature power spectrum}

We now turn to the scalar perturbations - basically density perturbations-, which develop on top of the FLRW metric in the different models 
considered\footnote{We disregarded vector and 
tensor perturbations in our study, whether they be intrinsic or extrinsic, e.g. those caused by CMB lensing through large-scale structure.}.
It is believed that the seeds of these perturbations are created, or at least enhanced during the inflationary epoch,
which provides the initial conditions, or primordial spectrum. Density perturbations evolve in any component which is able to ``clump'', i.e.
which can undergo some gravitational
instability; these are CDM, baryons and radiation. The interplay of these components shapes the power spectrum of the photons (CMB temperature spectrum), 
as well as the power
spectrum of the total matter content (matter power spectrum), discussed in the next subsection.
Before $z_{\rm rec}$, the baryon-photon fluid was tightly coupled and the antagonism between
gravitational pull of the CDM and baryons on the one hand, and radiation pressure of the photons on the other hand, created the pattern of
acoustic peaks seen in the CMB spectrum. While the amount of radiation in the universe can be determined, once we know the CMB temperature, the 
detailed form of the acoustic peaks depends upon the fraction of CDM and baryons, present in the universe.
As a result, the CMB spectrum allows to extract fundamental cosmological
parameters in a very robust\footnote{A priori, the density parameters at the present ($z=0$) may not coincide with those at $z_{\rm rec}$ probed by the CMB.
It is only that in the standard cosmological model, we know the exact inventory and therefore know how they evolve between then and now. Also, we stress that 
strictly speaking $\Omega_{\Lambda}$ can not be determined through the peaks alone, since a small curvature $\Omega_k$ is sufficient to mimic its effect.
The evidence for $\Omega_{\Lambda}$ through the CMB comes indirectly by allowing information from other observational sources.}
fashion: $\Omega_{m,0} h^2$ (refering to the total matter), $\Omega_{b,0} h^2$, $\Omega_{k,0}$, $\Omega_{\Lambda,0}$.

The density perturbations at the time of last scattering are imprinted on the CMB by 
variations of the temperature about the mean $\bar T_{\rm CMB}=2.726$ K. The size of this ``temperature contrast'' is of order $\delta T/\bar T_{\rm CMB} \approx 10^{-5}$. This observational fact, together with theoretical considerations, 
lend support to the idea that the evolution
of density perturbations can be treated in a perturbative manner, i.e. basically to linearize the original nonlinear coupled differential equations of motion of a multi-component cosmic
``fluid''. The evolution of perturbations is complicated by the expanding background, and the finite speed of light.
As a result, we are concerned with particle horizons of order $c/H(a)$ which change with time (or scale factor), and the size of a particular 
perturbation (i.e. its wavelength) needs to be compared with respect to this horizon: superhorizon perturbations have wavelengths larger than the horizon, 
while subhorizon perturbations have wavelengths smaller than the horizon. 

\begin{figure*} 
    \centering
     \includegraphics[width=12cm]{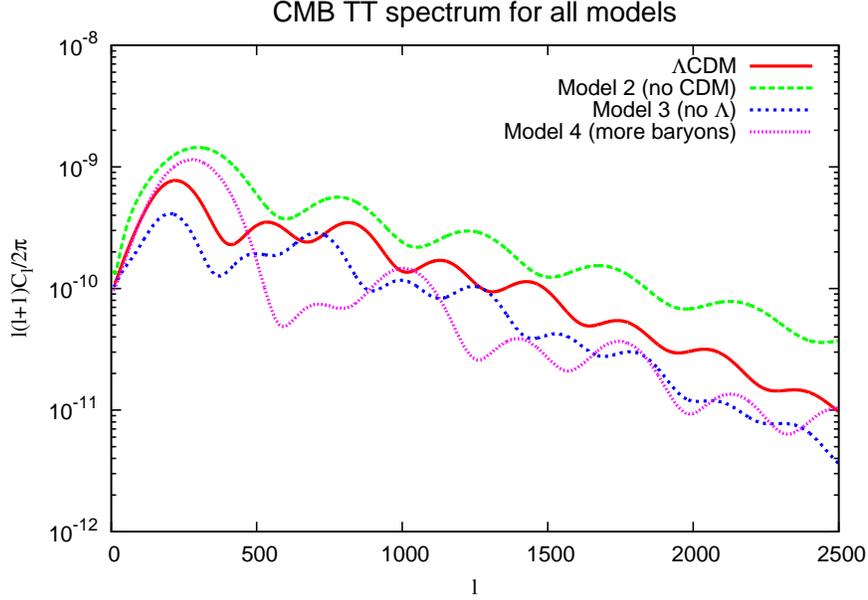}
 \caption{CMB temperature power spectrum (TT) as a function of mode number $l$ for all models considered.}
 \label{figureCMB}
\end{figure*}

\begin{figure*} 
    \centering
     \includegraphics[width=12cm]{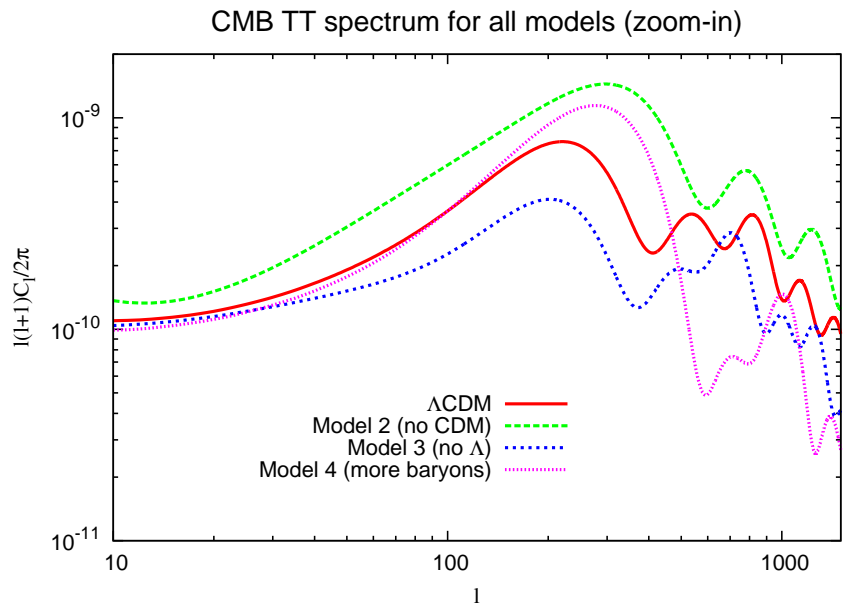}
 \caption{CMB temperature power spectrum (TT) as a function of mode number $l$ for all models considered, zoom-in to low $l$. Note the change to logarithmic scale
 on the $l$-axis.}
 \label{figureCMBzoom}
\end{figure*}

Now, let us introduce some basic notation.
In order to analyze the CMB temperature fluctuations, which are sourced by stochastic initial conditions in the density distribution, one adopts a statistical description: 
by introducing  
the Fourier decomposition of the density contrast, averaging the squares of the Fourier coefficients over different
realizations of the density field, one can calculate the variance as a measure of the clumpiness at different spatial or angular scales, respectively (see also the paragraph
surrounding equ.(\ref{delta})). On the celestial sphere, it is appropriate
to decompose the map of
 $\delta T/\bar T_{\rm CMB}$ as a function of position on the sky with spherical angles $(\theta, \phi)$ into spherical harmonics $Y_{lm}$,
 which are characterized by the mode number $l$ and order $m$:
 \begin{equation}
  \frac{\delta T}{\bar T_{\rm CMB}} (\theta, \phi) = \sum_{l=0}^{\infty} \sum_{m=-l}^{l} a_{lm} Y_{lm}(\theta,\phi),
 \end{equation}
 with the equivalent to Fourier coefficients 
 \begin{equation}
  a_{lm} = \int \frac{\delta T}{\bar T_{\rm CMB}}(\theta,\phi) Y_{lm}^* d\theta d\phi 
 \end{equation}
and the integral is over the whole sky.
 Because of assumed isotropy, the coefficients $a_{lm}$ are a function of $l$ only.
 The multipol moment $C_l$ is defined by 
 \begin{equation}
  C_l := \langle |a_{lm}|^2\rangle
 \end{equation}
where the brackets denote the average over $m$ for every $l$; $l$ equals a wavenumber on the sky and roughly $l \simeq \pi/\theta$ with the angular size $\theta$ on the sky.
The multipol moments are related to the autocorrelation function of the CMB temperature contrasts.
Different physical effects operate on different spatial scales, and these can be analyzed in a plot of $C_l$ versus $l$. It is 
customary\footnote{This way, the quantity $\frac{l(l+1)}{2\pi}C_l$ is constant for scale-invariant perturbations on large scales - a prediction from
inflationary models.} to plot the 
quantity $\frac{l(l+1)}{2\pi}C_l$ versus $l$; these are the CMB temperature power spectra plotted for all of our models:
Fig.\ref{figureCMB} shows plots over a range of $l=0-2500$, while Fig.\ref{figureCMBzoom} shows a zoom-in to low $l$ and a double-logarithmic representation, which is
commonly considered. In CLASS, the output CMB multipol moments $C_l$ are dimensionless.

Some important questions at hand with respect to the CMB power spectrum of the models can be summarized as follows:
\begin{enumerate}
\item Where is superhorizon and subhorizon physics at play?
\item Where is the location of the first acoustic peak; is it stable ? What is the height of the first acoustic peak; does it change ? As a result, what
does this imply for the robustness of determining the curvature of the model universe?
\item What is the location and height of the a) second peak, and b) third peak? How do they change?
\item How does the oscillation pattern at high mode numbers change, compared to $\Lambda$CDM? 
\end{enumerate} 
Naturally, the interpretation of the perturbation spectra is not a simple task, especially for newcomers to the field.  
However, all students were able to grasp the main
changes and their causes, which speaks for the usefulness of this lab exercise in teaching cosmology. Again, as this is no tutorial on CMB physics, we have to presume that the
reader is somewhat familiar with the standard CMB spectrum, and focus our attention mostly on the changes we can see in the exotic models. Also, our entire 
focus lies in the so-called
primary features/effects which are intrinsic to the CMB spectrum, so we do not consider secondary or tertiary effects, which shape the spectrum while the CMB photons
traverse the Universe up through the Milky Way to us.

The Hubble radius (or horizon) at $z_{\rm rec}$ represents an important scale: superhorizon perturbations have not yet entered the 
horizon at that time, and they preserve
the features of the primordial spectrum, which is nearly scale-free. Since the initial conditions are basically the same for each model 
(never mind some tiny variations in the exact values for $A_s$ and $n_s$),
all curves in Fig.\ref{figureCMBzoom} can be seen to become almost constant for low $l$: models with flat geometry roughly converge to the same value, as opposed to
Model 2 which has a positive curvature. 

On the other hand, subhorizon perturbations with wavelengths smaller than the horizon size were able to enter prior to $z_{\rm rec}$ and they are subject to
gravitational instability.
As time went by, perturbations with larger and larger wavelengths were able to enter the horizon (i.e. successively became subhorizon), 
until $z_{\rm rec}$ when the CMB was ``released''.  
The first acoustic peak corresponds to that oscillation whose size was just 
big enough to fill the horizon size at $z_{\rm rec}$, i.e. the location of the first peak is mainly determined by the global geometry of the Universe.
In flat universes, the first
peak is predicted to lie around $l \approx 220$. However, since a change of the baryon and CDM densities can affect the location of 
the first peak, as we will see,
we would need to fix those densities (corresponding to fixing the peak height) in order to receive a truly stable criterion for the determination of the global geometry.

The standard model $\Lambda$CDM is depicted as the red, solid curve in Fig.\ref{figureCMB}-\ref{figureCMBzoom}. We recognize the familiar run and height of the acoustic peaks,
as well as the strong damping tail due to photon diffusion for high $l$: the first peak is located around $l \simeq 220$, and the heights of the second and third peak are
roughly equal.
Now let us turn to Model 4 which, in a sense, is the closest to $\Lambda$CDM, although the highly increased baryon fraction makes a clear difference. Its spectrum is
given by the pink, dotted curve in Fig.\ref{figureCMB}-\ref{figureCMBzoom}. While Model 4 has a global flat geometry, we can clearly see that the first peak is shifted to higher $l$, 
compared to $\Lambda$CDM. This occurs because the increased baryon fraction implies a lower speed of sound of the coupled baryon-photon fluid, and as a result the sound
horizon shortens which shifts the peak towards higher $l$, away from the location of flat universes, similar to what a negative curvature would do.
Also, one main impact of varying the baryon fraction concerns the relative amplitudes of the peaks. Increasing the baryon fraction (as in Model 4)
leads to an increased baryon drag which, in turn, enhances the amplitudes of compressional (odd-numbered) peaks, while it suppresses rarefaction (even-numbered) peaks, 
thus the first and third relative peak heights are visibly enhanced, compared to $\Lambda$CDM. The reason can be found in the fact that a higher baryon fraction 
amounts to a higher mass of the baryon-photon fluid; the fluid is concentrated into denser and denser regions before radiation pressure is able to push back. 
Higher mass means lower frequency of the acoustic oscillations (akin to a simple harmonic oscillator), so the peaks occur on smaller spatial scales, i.e. higher $l$, compared to 
the lower baryon density of $\Lambda$CDM. Also, the energy of the oscillations increases which pushes the peak amplitudes to higher values. 
Furthermore, the diffusion damping by the photons is less efficient for higher baryon fraction, hence the power at high $l$ is relatively less 
suppressed in Model 4.

Now let us turn to Model 2 without CDM which is the green, long-dashed curve in Fig.\ref{figureCMB}-\ref{figureCMBzoom}. Here, we only have baryonic matter.
The growth of overdensities is severly limited here by essentially two factors, namely that decoupling of both, photons and baryons, occurs in the 
radiation-dominated epoch, when that universe expands faster than it would in a matter-dominated phase, and because there is no CDM. 
Hence, there are no CDM overdensities into which the baryon-photon fluid can fall, and overdensities in the baryons are severly suppressed.
It is thus very implausible that structure could have formed rapidly enough, in order
to give rise to the formation of the first galaxies in such models. Again, this is a classic problem of 
MOND universes\footnote{However, I should note that MOND models do not require an FLRW background metric, 
although they can be made consistent with FLRW. Also, the fundamental equations
of structure growth would be different. Although the impact of the modifications brought about
by MOND are believed to be small in the early Universe, matter density contrasts
can grow faster than $\propto a \sim t^{2/3}$ in a matter-dominated background.
In any case, if we want Model 2 to mimick a MOND universe, we should bear in mind that our interpretations are based upon important assumptions.}
and the timing issue of structure formation has been long considered
as one cornerstone supporting the evidence for dark matter. Yet, we want to understand the spectra in such a universe to compare them to the standard model.
Indeed, before the CMB spectrum was measured to the high precision we know it now (and before the high-$z$ supernovae distance measurements started to favour $\Lambda$CDM),
it was not so obvious to dismiss a low total matter content. A discussion of CDM versus MOND (resp. low total matter) models and their impact onto the CMB spectrum can
be found e.g. in \cite{Mc}, which also served the author as a source to understand the outcome of Model 2. The CMB spectrum of Model 2 looks like a very 
nice harmonic, damped oscillation, which shows the baryon-photon fluid at play, completely unhindered by CDM. 
However, in order to connect to what we have said previously, we notice that in the limit where baryons constitute all of the matter, the even-numbered
peaks are suppressed to the point of disappearing! Therefore, what looks like the second and third peak, is actually the third and fifth peak, respectively. 
Thus, the typical ``odd-even effect'' as a measure of constraining the fraction of CDM versus baryons has disappeared altogether.
The peaks are perceptibly broader and the amplitudes are higher in Model 2, compared to all the other models. Also, as seen already in Model 4, 
the location of the first peak shifts
to higher $l$ (smaller angular scales). 
Moreover, in this model we have the smallest amount of suppression of power due to photon diffusion at high $l$.

Finally, we turn to Model 3 without $\Lambda$, which is strongly CDM-dominated. It is depicted by the blue, short-dashed curve in Fig.\ref{figureCMB}-\ref{figureCMBzoom}.
Increasing the total matter content in such a dramatic way enforces a suppression of the entire
spectrum and its peak amplitudes: the first two peaks become suppressed by a larger factor, enhancing the relative height of the third and following peaks. 
The well-known fact that the height of the third peak is a particularly good indicator for $\Omega_{\rm cdm}$ can be clearly appreciated in this model.
Although the total baryonic content is the same as in $\Lambda$CDM, its comparatively diminished impact causes an overall shift of
the acoustic peaks towards lower $l$ (larger angular scales), as well as an increase in the frequency of the oscillations.

\subsection{Matter power spectrum}

In this subsection, we turn to the (total) matter power spectrum of our models. We already indicated briefly the statistical description of density perturbations 
before we introduced the CMB multipol moments. Let us now introduce the notation we use here: using the matter density contrast at a fixed time $t$ (or $a$ or $z$),
$\delta(\vec{x}) = (\rho(\vec{x})-\bar \rho)/\bar \rho$,
where $\bar \rho$ shall denote the homogeneous background matter density, and its Fourier decomposition (assuming $\delta(\vec{x})$ is a periodic function over a side-length $L$
of the ``universe-box''),
\begin{equation} \label{delta}
 \delta(\vec{x}) = \sum_k \delta_k e^{i\vec{k}\cdot \vec{x}},
\end{equation}
where $\delta_k$ are the Fourier coefficients and wavevector $\vec{k} = (k_1,k_2,k_3), k_i = 2\pi n_i/L, n_i \in \mathbb{N}$, we can again estimate the variance by averaging
the squares of the Fourier coefficients over different realizations of the density field for a fixed wavenumber $k := |\vec{k}|$;
this is the power spectrum
\begin{equation}
 P(k) = \langle |\delta_k|^2\rangle
\end{equation}
and because of isotropy, $P(k)$ does not depend on the direction of $\vec{k}$. By analogy to the $C_l$'s, the power spectrum $P(k)$ is related to the autocorrelation function of the 
density contrast. 

\begin{figure*} 
    \centering
     \includegraphics[width=12cm]{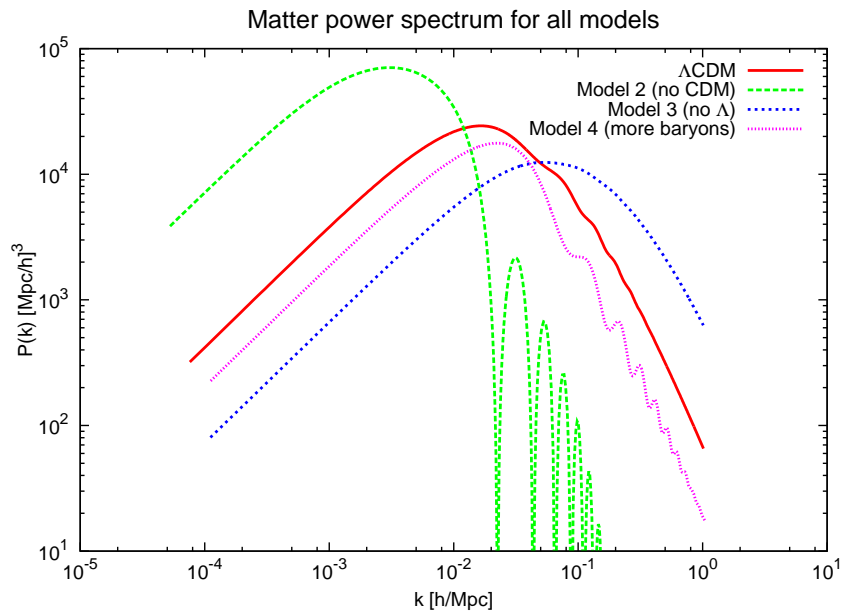}
 \caption{Matter power spectrum as a function of wavenumber $k$ for all models considered.}
 \label{figure3}
\end{figure*}

We plot $P(k)$ at $z=0$ in Fig.\ref{figure3} as a function of wavenumber $k$, for all models considered. In the standard cosmological scenario, the matter power spectrum
can be accurately described by linear theory up to about $k \sim 0.1$ h Mpc$^{-1}$. In principle, one can estimate the nonlinear contribution by routines like ``halofits'', and
CLASS is also able to do this (it is taken into account in $\Lambda$CDM (2018), but the resulting difference is of no concern to our study here; see also the Appendix).

Some important questions at hand with respect to the matter power spectrum of the models can be summarized as follows:
\begin{enumerate}
\item Where is superhorizon and subhorizon physics at play?
\item What is the overall shape? Are there changes? 
\item Does the peak shift location or change height?
\item How does the oscillation pattern at large $k$ (i.e. small spatial scales) change, compared to $\Lambda$CDM? 
\end{enumerate} 

In the last subsection, we have already identified the main physical causes of the changes we see between models. Of course, the same physics is at play 
in the matter power spectrum. Therefore, it simply remains to highlight the visible changes we see in Fig.\ref{figure3}.

In a sense, the CMB and matter power spectra look ``similar'' in that there is a rise to a first peak, followed by smaller peaks with less power.
While the CMB spectrum reflects much more visibly the physics of the coupled baryon-photon fluid, the matter power spectrum is mostly shaped by CDM which suppresses
these peaks in $P(k)$. This picture is nicely illustrated in the standard model, which is the red, solid curve in Fig.\ref{figure3}: while the superhorizon perturbations
are well described by a nearly scale-free run, $P(k) \propto k^{n_s}$ with $n_s \lesssim 1$, we observe a falloff above around $k \sim 10^{-2}$ h Mpc$^{-1}$, which indicates
the transition to the subhorizon regime where the acoustic oscillations of the baryon-photon fluid make their mark as small ``wiggles''. Here, in the matter power spectrum, 
these oscillations are referred to as baryonic acoustic oscillations (BAO). A discussion of BAOs in models of varying baryon and total matter content can be found
in \cite{EH}.

The overall slope in the subhorizon regime goes as $P(k) \sim k^{-3}$, as expected.
Let us consider again the effect of increasing the baryon fraction by looking first at Model 4 - the ``closest'' to $\Lambda$CDM-, 
which is the pink, dotted curve in Fig.\ref{figure3}. While the overall slopes for small and large $k$ do not change, the BAO wiggles have markedly increased, indeed.
We also see an overall suppression of the entire spectrum, another feature of elevating the baryon fraction.
Now, let us turn to Model 2, depicted as the green, long-dashed curve. This is a pure baryon power spectrum, and it looks like expected:
the amplitude of the first peak is enhanced, compared to $\Lambda$CDM, similar to what we see in Fig.\ref{figureCMBzoom}, and the oscillations are much more pronounced than in 
any other model. Yet, their power is severly suppressed because this is a model of low total matter content without CDM. As Model 2 describes a positively curved universe,
we see more power at small $k$, but the superhorizon slope is unchanged.
Finally, Model 3 without $\Lambda$ is strongly CDM-dominated and therefore the BAO feature is highly suppressed to a level where it is invisible in this plot, see the blue,
short-dashed curve in Fig.\ref{figure3}.
It is a very smooth power spectrum which is almost totally shaped by CDM alone and, again, it preserves the overall slopes for small and large $k$.

A characteristic scale of importance is the horizon size at matter-radiation equality, as density perturbations behave differently in a radiation-dominated universe versus a 
matter-dominated one.
Comparing the location of the maximum (the ``first peak''), we can clearly see that the peak shifts to the right, i.e. to larger $k$ 
(smaller spatial scales), if matter-radiation equality happens earlier, i.e. for higher $z_{\rm eq}$. For $\Lambda$CDM (2018) with $z_{\rm eq} = 3407$, the maximum is at 
$k_H = 1.6576\times 10^{-2}$ h Mpc$^{-1}$ which corresponds to a spatial scale of $379$ h$^{-1}$ Mpc: most of the variance in the cosmic density field in the universe
at the present epoch is below that scale. By the same token, the two extremes of Model 2 ($z_{\rm eq} = 526$) and Model 3 ($z_{\rm eq} = 10908$) have their maximum
power at wavenumbers of $k_H = 3.1053\times 10^{-3}$ h Mpc$^{-1}$ and $k_H = 5.3174\times 10^{-2}$ h Mpc$^{-1}$, respectively, which corresponds to scales of 
$2023$ h$^{-1}$ Mpc and $118$ h$^{-1}$ Mpc, respectively. Like $\Lambda$CDM, Model 4 with $z_{\rm eq} = 5745$ lies in between these extremes, 
$k_H = 2.2244\times 10^{-2}$ h Mpc$^{-1}$ corresponding to
$282$ h$^{-1}$ Mpc.

\section{Summary}
\label{sec:summary}
We reported on a computer lab exercise, as part of an advanced cosmology lecture course of the author. We used the open source software CLASS, in order to
calculate the background evolution, the CMB temperature power spectrum and the matter power spectrum for the standard cosmological model $\Lambda$CDM, as well as for three
exotic models with very different cosmological density parameters, compared to $\Lambda$CDM. Our chosen exotic models are firmly ruled out by modern cosmological and 
astronomical observations. However, the purpose of our exercise was not only to learn to use CLASS, but mostly to gain intuition and understanding of the impacts of baryons, CDM
and $\Lambda$ on the CMB spectra. By comparing these impacts among the different models, the students were able in a first-hand approach to grasp the 
importance and prevalence of the current cosmological standard model $\Lambda$CDM in the interpretation of modern data.

\ack

The author acknowledges support by the Austrian Science Fund FWF through an Elise Richter fellowship, grant nr. V 656-N28.

\appendix

\section{$\Lambda$CDM models}

During the early days of establishing the current standard model, $\Lambda$CDM was traditionally characterized 
by ``round'' numbers, $\Omega_{\rm{cdm},0} = 0.25, \Omega_{b,0} = 0.05$ and $\Omega_{\Lambda,0} = 0.70$, which sum up to give a flat global geometry.
But what constitutes \textit{the} $\Lambda$CDM benchmark (or best-fit) model
is nowadays determined by ever more precise measurements of different cosmological observables, and it is fair to say that the advent of modern CMB observations
has helped to shape and to strengthen what constitutes now $\Lambda$CDM. Nevertheless, we can say that $\Lambda$CDM will always be characterized by the above rough share between 
$\Lambda$ and CDM, with only a small amount of baryons.  Photons and neutrinos are so much subdominant at the present time, that they do not even receive explicit mention. However,
the era of modern precision cosmology has brought about the necessity to include the mass of the neutrinos, as well as other factors previously disregarded 
(e.g. lensing of the CMB photons on their way to us due to large-scale structure). 

In our CLASS lab exercise, we considered an old $\Lambda$CDM model from the early time of CLASS (termed $\Lambda$CDM (2011) with input file
\verb"lcdm.ini"), as well as a new model which comes with 
CLASS release 2.7.1 (termed $\Lambda$CDM (2018) with input file \verb"base_2018_plikHM_TTTEEE_lowl_lowE_lensing.ini") which reflects the most 
current Planck data release of 2018 (see also Sec.\ref{sec:Intro}). $\Lambda$CDM (2018) includes one massive neutrino species and takes care of CMB lensing. 
Many more parameters of $\Lambda$CDM (2018) are updated with respect to older benchmark models, like Helium fraction, 
or the amplitude and index of the primordial scalar spectrum, $A_s$ and $n_s$, respectively. Also, it includes an estimate of the nonlinear contribution to
$P(k)$ and $C_l$, using a 'halofit' routine.   

\begin{figure*} 
\begin{minipage}{0.5\linewidth}
     \centering
     \includegraphics[width=8cm]{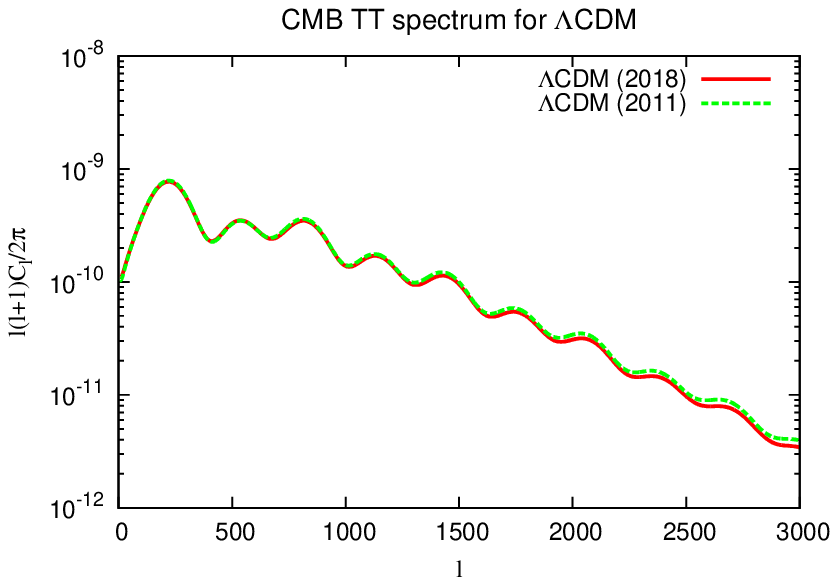}
     \vspace{0.05cm}
    \end{minipage}
    \begin{minipage}{0.5\linewidth}
      \centering\includegraphics[width=8cm]{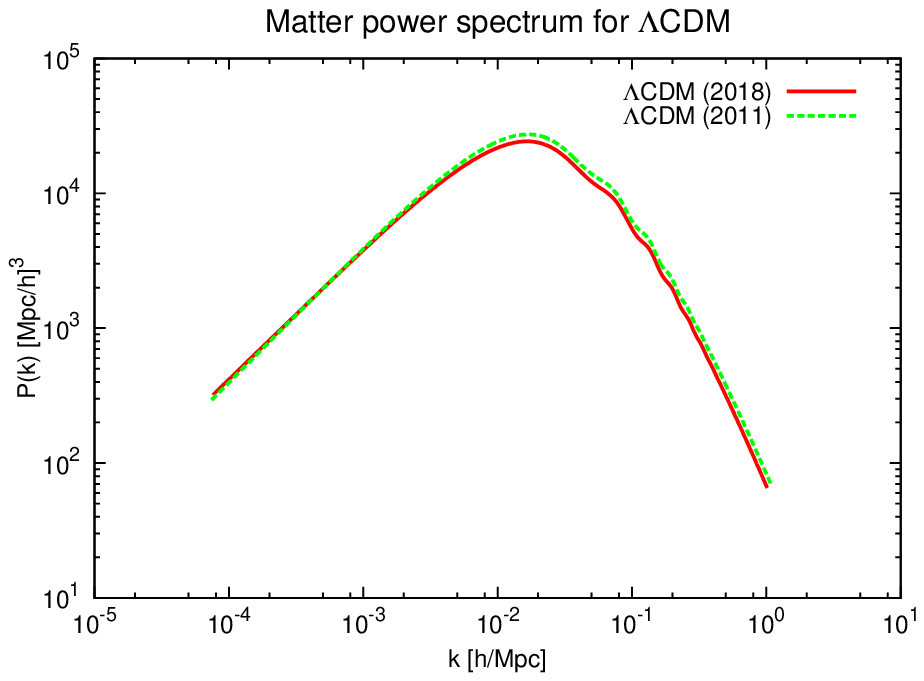}
     \hspace{0.05cm}
    \end{minipage}
 \caption{Comparison of the (unlensed) CMB TT spectrum of $\Lambda$CDM (2011) with the (lensed) CMB TT spectrum of $\Lambda$CDM (2018) (left) and comparison of
 the respective matter power spectra (right). More background can be found in the text.}
 \label{lcdm}
\end{figure*}

Of course, the deviations between $\Lambda$CDM (2011) and (2018) are marginal, compared to the exotic models considered for this study, so from this perspective it does not matter
which reference $\Lambda$CDM model we choose. For the sake of this article, we pick $\Lambda$CDM (2018) in all of our plots. In order to demonstrate their similarity, 
compared to the other models, we show plots of the CMB temperature and matter power spectra for both $\Lambda$CDM models in Fig.\ref{lcdm}.

\section*{References}

\begin{thebibliography}{99}


\bibitem{S1}
S.Alam, M.Ata, S.Bailey et al,
\emph{The clustering of galaxies in the completed SDSS-III Baryon Oscillation Spectroscopic Survey: cosmological analysis of the DR12 galaxy sample}, MNRAS, \textbf{470}, Issue 3, p.2617-2652 (2017)

\bibitem{S2}
M.Blomqvist, H. du Mas des Bourboux, N.G.Busca et al,
\emph{Baryon acoustic oscillations from the cross-correlation of Ly-alpha
absorption and quasars in eBOSS DR14}, A\& A, \textbf{629}, A86, 18pp. (2019)

\bibitem{CMB1}
Planck Collaboration: N.Aghanim, Y.Akrami, M.Ashdown et al,
\emph{Planck 2018 results.VI. Cosmological parameters},
arXiv: 1807.06209

\bibitem{CMB2}
G.Simard, Y.Omori, K.Aylor et al, 
\emph{Constraints on Cosmological Parameters from the Angular Power Spectrum of a Combined 2500 deg2 SPT-SZ and Planck Gravitational Lensing Map}, 
ApJ, \textbf{860}, Issue 2, article id. 137, 9 pp. (2018)

\bibitem{D1}
DES Collaboration: T.M.C.Abbott, S.Allam, P.Andersen et al,
\emph{First Cosmology Results using Type Ia Supernovae from the Dark Energy Survey: Constraints on Cosmological Parameters}, 
ApJL, \textbf{872}, Issue 2, article id. L30, 9 pp. (2019)

\bibitem{D2}
A.G.Riess, S.Casertano, W.Yuan et al,
\emph{Milky Way Cepheid Standards for Measuring Cosmic Distances and Application to Gaia DR2: Implications for the Hubble Constant},
ApJ, \textbf{861}, Issue 2, article id. 126, 13 pp. (2018)

\bibitem{D3}
A.G.Riess, S.Casertano, W.Yuan, L.M.Macri, D.Scolnic,
\emph{Large Magellanic Cloud Cepheid Standards Provide a 1\% Foundation for the Determination of the Hubble Constant and Stronger Evidence for Physics beyond $\Lambda$CDM},
ApJ, \textbf{876}, Issue 1, article id. 85, 13 pp. (2019)

\bibitem{D4}
C.R.Burns, E.Parent, M.M.Phillips et al,
\emph{The Carnegie Supernova Project: Absolute Calibration and the Hubble Constant},
ApJ, \textbf{869}, Issue 1, article id. 56, 23 pp. (2018)


\bibitem{HS95}
W.Hu, N.Sugiyama, \emph{Anisotropies in the Cosmic Microwave Background: an Analytic Approach}, 
ApJ, \textbf{444}, 489, 18 pp. (1995)


\bibitem{HS96}
W.Hu, N.Sugiyama, \emph{Small-Scale Cosmological Perturbations: an Analytic Approach}, 
ApJ, \textbf{471}, 542-570 (1996)


\bibitem{Mukh}
V.Mukhanov, \emph{``CMB-Slow'' or How to Determine Cosmological Parameters by Hand?}, 
Intern.J.Theor.Phys., \textbf{43}, issue 3, 623-668 (2004)


\bibitem{Sugi}
N.Sugiyama, \emph{Introduction to temperature anisotropies of Cosmic Microwave Background radiation}, 
Prog.Theor.Exp.Phys., \textbf{2014}, Issue 6, 13 pp. (2014)


\bibitem{Staggs}
S.Staggs, J.Dunkley, L.Page, \emph{Recent discoveries from the cosmic microwave background: a review of recent progress},
Rep.Progr.Phys., \textbf{81}, Nr.4, 044901 (2018)


\bibitem{CLASS}
J.Lesgourgues, \emph{The Cosmic Linear Anisotropy Solving System (CLASS) I: Overview}, arXiv:1104.2932


\bibitem{CLASS2}
D.Blas, J.Lesgourgues, T.Tram,
\emph{The Cosmic Linear Anisotropy Solving System
(CLASS) II: Approximation schemes}, JCAP 07 (2011) 034


\bibitem{CLASS3}
J.Lesgourgues,
\emph{The Cosmic Linear Anisotropy Solving System
(CLASS) III: Comparison with CAMB for $\Lambda$CDM}, arXiv:1104.2934


\bibitem{CLASS4}
J.Lesgourgues, T.Tram,
\emph{The Cosmic Linear Anisotropy Solving System
(CLASS) IV: Efficient implementation of non-cold
relics}, JCAP 09 (2011) 032


\bibitem{Mc}
S.S.McGaugh, \emph{Distinguishing between CDM and MOND: Predictions for the Microwave Background}, 
ApJ, \textbf{523}, L99 (1999)
  
  
\bibitem{EH}
D.J.Eisenstein, W.Hu, \emph{Baryonic Features in the Matter Transfer Function}, ApJ, \textbf{496}, 605 (1998)

\end{thebibliography}
\end{document}